\documentstyle[11pt]{article}

\catcode`\@=11
\ifcase \@ptsize
\or 
\or 
\fi
\advance\textheight by \topskip
\catcode`\@=12

\newcommand{\beqn}{\begin{eqnarray}}
\newcommand{\eeqn}{\end{eqnarray}}

\newcommand{\bbar}[1]{\mbox{$\bar #1$}}

\newcommand{\Hrel}{H_{\mbox{\scriptsize rel}}}
\newcommand{\Lrel}{L_{\mbox{\scriptsize rel}}}

\newcommand{\rmd}{\mbox{d}}
\newcommand{\rme}{\mbox{e}}
\newcommand{\rmi}{\mbox{i}}

\newcommand{\srmi}{\mbox{\scriptsize i}}

\newcommand{\dd}[2]{{\rmd{#1}\over\rmd{#2}}}
\newcommand{\pdd}[2]{{\partial{#1}\over\partial{#2}}}

\newcommand{\wt}{\widetilde}

\catcode`\@=11
\newcommand{\ccite}[1]
{\@ifundefined{b@#1}{\bf ?}{\@nameuse{b@#1}}}
\catcode`\@=12

\begin{document}

\begin{titlepage}

\vspace*{\fill}

\centerline{\LARGE\bf
Solution of the Three--Anyon Problem
}

\vspace{2cm}


\centerline{\large Stefan Mashkevich}
\centerline{\large N.N.\ Bogolyubov Institute for Theoretical Physics}
\centerline{\large 252143 Kiev, Ukraine}

\vspace{0.5cm}

\centerline{\large Jan Myrheim, K{\aa}re Olaussen and Ronald Rietman}
\centerline{\large Institutt for fysikk, NTH}
\centerline{\large N--7034 Trondheim, Norway}

\vspace{0.5cm}

\centerline{\today}

\vspace{2.5cm}

\centerline{\bf Abstract}

\vspace{0.4cm}

We solve, by separation of variables,
the problem of three anyons with a harmonic oscillator
potential. The anyonic symmetry conditions from cyclic
permutations are separable in our coordinates.
The conditions from two-particle transpositions
are not separable, but can be expressed as reflection
symmetry conditions on the wave function and its
normal derivative on the boundary of a circle.
Thus the problem becomes one-dimensional.
We solve this problem numerically by discretization.
$N$-point discretization with very small $N$ is
often a good first approximation, on the other hand
convergence as $N\to\infty$ is sometimes very slow.

\vspace*{\fill}

\end{titlepage}

\section{Introduction}

Anyons are two-dimensional point particles
with fractional statistics
\nocite{LeinaasMyrheim77}
\nocite{Goldinetal80}
\nocite{Goldinetal81}
\nocite{Wilczek82}
\nocite{Wilczek82b}
{[\ccite{LeinaasMyrheim77}--%
\ccite{Wilczek82b}]}, which seem to be useful
as models for the quasiparticle excitations
in the fractional quantum Hall systems.
For reviews see e.g.\
\nocite{Wilczek:FractStat}
\nocite{Forte92}
\nocite{IengoLechner92}
\nocite{Lerda:Anyons}
\nocite{Stone:QHE}
{[\ccite{Wilczek:FractStat}--\ccite{Stone:QHE}]}.

We use here what is known as the ``anyonic''
gauge, in which the vector potential associated
with particle statistics vanishes, and the wave
functions are many-valued, as follows. Go along
a continuous curve in the $N$-anyon configuration
space, starting and ending with the same
configuration $A$, but such that two particles are
interchanged in the counterclockwise direction,
while no other particles are encircled or
interchanged. Then the value of a wave
function $\psi$ must change continuously from
$\psi(A)$ to $\rme^{\srmi\theta}\,\psi(A)$.
The phase angle $\theta$ is a constant,
independent of the configuration $A$ and
the wave function $\psi$. It defines
the statistics of the particles, it is
$0$ for bosons, $\pi$ for fermions, and
may take any real value for anyons.

The energy spectrum for two anyons in
a harmonic oscillator potential is easily
found, and it interpolates continuously
between the boson and fermion spectra
when $\theta$ varies
\cite{LeinaasMyrheim77}.
The energies vary linearly with $\theta$,
sometimes with a discontinuous
change in slope at the Bose points.
A more general class of quadratic
Hamiltonians, including the case of
a constant magnetic field, can be treated
just as easily.

The same problem for more than two anyons
is much more difficult. Wu found a class
of exact solutions that generalize
the two-anyon solutions \cite{Wu84a}.
More general exact solutions have been found,
but all have energies
that depend linearly on $\theta$
\nocite{JohnsonCanright90}
\nocite{Polychronakos91}
\nocite{Chou91}
\nocite{Chou91a}
\nocite{Chou91b}
\nocite{Dunneetal91}
\nocite{Dunneetal91a}
\nocite{Dunneetal92}
\nocite{Grundbergetal91a}
\nocite{KarlhedeWesterberg92}
\nocite{Sen92}
\nocite{Sen92a}
\nocite{RuudRavndal92}
\nocite{ChoRim92}
\nocite{Choetal92}
\nocite{Mashkevich92}
\nocite{Mashkevich92a}
\nocite{Mashkevich93}
\nocite{Roy93}
\nocite{Dateetal94}
{[\ccite{JohnsonCanright90}--%
\ccite{Dateetal94},\ccite{Lerda:Anyons}]}.
Thus, even the three-anyon ground state
close to Fermi statistics, which has
an energy varying non-linearly with $\theta$,
is still not known analytically.
Even though the exact $\theta$-dependence
is unknown for many energy levels,
the spectral flow with $\theta$ from the boson
to the fermion spectrum is known
\cite{Mashkevich92a,Mashkevich93}.

Sen found an interesting ``supersymmetry''
in the three-anyon spectrum, excluding
some of the states with linearly varying
energy \cite{Sen92,Sen92a}. His supersymmetry
transformation transforms bosons into fermions
and vice versa, and more generally transforms
$\theta$ into $\pi-\theta$.

The lowest part of the complete energy spectrum
of three or four anyons in a harmonic
oscillator potential has been calculated
numerically
\nocite{Sporreetal91}
\nocite{Murthyetal91}
\nocite{Sporreetal92}
\nocite{Lawetal92}
\nocite{Lawetal94}
{[\ccite{Sporreetal91}--\ccite{Lawetal94}]}.
In the three-anyon problem, good analytical
approximations to the wave functions
corresponding to non-linear variation of
energy are known \cite{ChinHu92}.
More systematic approximation methods
are perturbation theory, starting from
the known boson and fermion spectra
\nocite{KhareMcCabe91}
\nocite{Chouetal92}
\nocite{AmelinoCamelia93}
{[\ccite{KhareMcCabe91}--%
\ccite{AmelinoCamelia93}]}, and
the Hartree--Fock approximation
\cite{Hannaetal88,Sitko92}.

One reason for the interest in the spectra
of two- and three-particle systems in
an external harmonic oscillator potential
is that they can be used for calculating
the second and third virial coefficients
in the equation of state for an ideal gas
of free particles, i.e.\ when there is
no interaction apart from the statistics
interaction. We will not discuss this
problem here, but refer to the literature
\nocite{Arovasetal85}
\nocite{VeigyOuvry92a}
\nocite{VeigyOuvry93a}
\nocite{Sporreetal93}
\nocite{EmparanValle93}
\nocite{MyrheimOlaussen93}
{[\ccite{Arovasetal85}--%
\ccite{MyrheimOlaussen93},%
\ccite{JohnsonCanright90},%
\ccite{Sen92},\ccite{Sen92a},%
\ccite{Lawetal92},%
\ccite{Lawetal94}]}.

We show here how to solve the three-anyon problem
with a harmonic oscillator potential, by separation
of variables in a suitable set of coordinates.
The same method applies to a class of similar
problems, as mentioned above.
An anticlockwise cyclic interchange of the three
particles gives a phase factor of
$\rme^{2\srmi\theta}$ in the wave function, and
this condition is compatible with the separation
of variables. However, the interchange of only
two particles gives a condition which can in general
only be satisfied by a superposition of separated
wave functions. Thus the separation is incomplete.
We find general solutions numerically, but we hope
that this method may also lead to some progress
in the search for analytical solutions.

\section{Formulation of the Problem}

It is a well known procedure to describe a particle
in the plane by a complex coordinate $z$ and its
complex conjugate $\bbar{z}$, which may be scaled
so as to become dimensionless. In the three-anyon
problem the key to the separation of variables is
a transformation from the particle coordinates
$z_1,z_2,z_3$ to the centre of mass coordinate $Z$
and the relative coordinates $u,v$, defined by
\beqn
\label{eq:defZuv}
Z&\!\!\!=&\!\!\!{1\over\sqrt{3}}\left(
z_1+z_2+z_3\right),\nonumber\\
u&\!\!\!=&\!\!\!{1\over\sqrt{3}}\left(
z_1+\eta z_2+\eta^2 z_3\right),\\
v&\!\!\!=&\!\!\!{1\over\sqrt{3}}\left(
z_1+\eta^2 z_2+\eta z_3\right).\nonumber
\eeqn
Here $\eta=\exp(2\rmi\pi/3)=(-1+\rmi\sqrt{3})/2$
is a cube root of unity, with
$\eta^2=\eta^{-1}=\bbar{\eta}$ and
$1+\eta+\eta^2=0$. Hence,
\beqn
u&\!\!\!=&\!\!\!{1\over\sqrt{3}}\left(
\eta(z_2-z_1)+\eta^2(z_3-z_1)\right),
\nonumber\\
v&\!\!\!=&\!\!\!{1\over\sqrt{3}}\left(
\eta^2(z_2-z_1)+\eta(z_3-z_1)\right).
\eeqn

This coordinate transformation is
a discrete Fourier transformation, which
means that a cyclic interchange of particle
positions,
\beqn
\label{eq:3cycle}
(z_1,z_2,z_3)\mapsto
(\wt{z}_1,\wt{z}_2,\wt{z}_3)=
(z_2,z_3,z_1)\;,
\eeqn
becomes diagonal,
\beqn
(Z,u,v)\mapsto(\wt{Z},\wt{u},\wt{v})=
(Z,\eta^2 u,\eta v)\;.
\eeqn
The interchange of particles 2 and 3
is just an interchange of $u$ and $v$.
Note that these two permutations generate
the whole symmetric group $S_3$.
A similar treatment of permutation symmetry is
known in nuclear physics
\cite{AguileraNavarroetal68,Jacksonetal70}
(we thank Alex Lande for pointing this out).

Three particles in the plane define a triangle.
The ratio $s=(z_3-z_1)/(z_2-z_1)$ is real
when the triangle is degenerate so that
the particles lie on a straight line.
We define the orientation of a non-degenerate
triangle as positive or negative depending on
whether the imaginary part of $s$ is positive
or negative. We have that
\beqn
{|u|\over|v|}=
{|\eta+\eta^2 s|\over|\eta^2+\eta s|}=
{|s+\eta^2|\over|s+\eta|}\;.
\eeqn
Hence $|u|=|v|$ if $s$ is real,
$|u|<|v|$ if the orientation
of the triangle is positive, and $|u|>|v|$
if the orientation is negative.

The quantization of the centre of mass motion is
trivial, and the problem we want to discuss here
is the simultaneous diagonalization of
the dimensionless relative Hamiltonian $\Hrel$ and
relative angular momentum $\Lrel$, defined by
\beqn
\Hrel&\!\!\!=&\!\!\!
-\pdd{^2}{u\,\partial\bbar{u}}
-\pdd{^2}{v\,\partial\bbar{v}}
+u\bbar{u}+v\bbar{v}\;,
\nonumber\\
\Lrel&\!\!\!=&\!\!\!
u\,\pdd{}{u}+v\,\pdd{}{v}
-\bbar{u}\,\pdd{}{\bbar{u}}
-\bbar{v}\,\pdd{}{\bbar{v}}\;.
\nonumber
\eeqn

\section{Separation of Variables}

The three-particle configuration is completely
described by a total scale factor $r>0$,
a relative scale factor $q\geq 0$, and two angles
$\varphi_1$ and $\varphi_2$ such that
\beqn
u={rq\,\rme^{\,\srmi\varphi_1}
\over\sqrt{2(1+q^2)}}\;,\qquad
v={r\,\rme^{\,\srmi\varphi_2}
\over\sqrt{2(1+q^2)}}\;.
\eeqn
These are the hyperspherical coordinates of
Kilpatrick and Larsen \cite{KilpatrickLarsen87},
except that they use $\varphi_1\pm\varphi_2$
instead of $\varphi_1$ and $\varphi_2$.
We now have that
\beqn
\Hrel=
-{1\over 2r^3}\,\pdd{}{r}\,r^3\,\pdd{}{r}
-{1+q^2\over 2r^2}\left(
{1+q^2\over q}\pdd{}{q}\,q\,\pdd{}{q}+
{1\over q^2}\,\pdd{^2}{{\varphi_1}^2}+
\pdd{^2}{{\varphi_2}^2}\right)
+{r^2\over 2}\;,
\eeqn
and
\beqn
\Lrel=-\rmi\left(
\pdd{}{\varphi_1}+\pdd{}{\varphi_2}\right).
\eeqn

Assume that the wave function of the relative
motion is
\beqn
\psi=\psi(r,q,\varphi_1,\varphi_2)=
f(r)\,g(q)\,\rme^{\,\srmi\,(j\varphi_1+k\varphi_2)}\;.
\eeqn
Then the eigenvalue equation $\Hrel\,\psi=E\psi$
separates into an angular eigenvalue equation,
\beqn
\label{eq:eigvalang}
(1+q^2)\left(
-{1+q^2\over q}\dd{}{q}\,q\,\dd{}{q}+
{j^2\over q^2}+k^2\right)g=\lambda g\;,
\eeqn
with $\lambda$ as eigenvalue, and a radial
equation,
\beqn
-{1\over 2}\,f''(r)-{3\over 2r}\,f'(r)
+\left({\lambda\over 2r^2}+{r^2\over 2}\right)
f(r)=Ef(r)\;.
\eeqn
A general wave function can be written as
a linear combination of such separated wave
functions. As shown in the next section, we
need linear combinations, where $\lambda$
and $j+k$ are constant but $j-k$ varies, in order
to satisfy the anyonic boundary conditions.

In the usual way we find the solutions
of the radial equation to be
\beqn
f(r)=r^{\mu}\,\rme^{-r^2/2}
\sum_{m=0}^{n_r} a_m\,r^{2m}\;,
\eeqn
with $n_r=0,1,2,\ldots$, with $\mu>-2$ for
normalizability, $\lambda=\mu(\mu+2)$, and
\beqn
E=2+\mu+2n_r\;.
\eeqn
The coefficients in the polynomial satisfy
the recursion formula
\beqn
2(m+1)(m+2+\mu)\,a_{m+1}=(2m+2+\mu-E)\,a_m\;.
\eeqn

The angular equation has two asymptotic solutions
$q^{\pm j}$ in the limit $q\to 0$. We exclude
the singular solution (for $j=0$ the singularity is
logarithmic). In fact there is no reason for
any singularity at $q=0$, where
the configuration is an equilateral triangle.
This leaves a solution which is unique up to
a multiplicative constant, and with a convenient
normalization it may be written as
\beqn
\label{eq:gqdef}
g(q)=q^{|j|}\,(1+q^2)^{\kappa}\,F(a,b;c;-q^2)\;.
\eeqn
The constant $\kappa$ here may be chosen in
one of two ways,
\beqn
\kappa={\mu\over 2}+1\qquad\mbox{or}\qquad
\kappa=-{\mu\over 2}\;,
\eeqn
giving two different representations of
the same solution. The constants
\beqn
a={|j|+|k|\over 2}+\kappa\;,\qquad
b={|j|-|k|\over 2}+\kappa\;,\qquad
c=1+|j|\;,
\eeqn
define the hypergeometric series
\beqn
F(a,b;c;x)=
\sum_{m=0}^{\infty}
{(a)_m\,(b)_m\over(c)_m}\,
{x^m\over m!}\;,
\eeqn
where, e.g., $(a)_0=1$,
$(a)_{n+1}=(a)_n\,(a+n)$.
The convergence radius for this series
is 1. There exist however other
representations,
\beqn
g(q)&\!\!\!=&\!\!\!
q^{|j|}\,(1+q^2)^{\kappa-a}\,
F\!\left(a,c-b;c;{q^2\over 1+q^2}\right)
\nonumber\\
&\!\!\!=&\!\!\!q^{|j|}\,(1+q^2)^{\kappa-b}\,
F\!\left(c-a,b;c;{q^2\over 1+q^2}\right)\;,
\eeqn
where the series converge for all real $q$.

A useful parameter, as will be seen below,
is the logarithmic derivative of $g$ at $q=1$,
\beqn
\beta={g'(1)\over g(1)}=|j|+\kappa
-{2ab\,F(a+1,b+1;c+1;-1)\over c\,F(a,b;c;-1)}\;.
\eeqn

\section{The Anyonic Boundary Conditions}

For three identical particles there is a six-fold
identification of points in the Euclidean relative
space. We will restrict the wave functions to
the region $0\leq q\leq 1$, which corresponds to
all the positively oriented triangles, but is still
a three-fold covering of the true configuration space.
The boundary conditions defining the particles
to be anyons are of two types, since there are
two classes of non-trivial permutations. The first
class contains the three-particle cyclic permutations,
which leave $q$ invariant. The second class contains
the two-particle interchanges, which transform $q$
into $1/q$, and so give boundary conditions at $q=1$.

Consider first a continuous, counterclockwise and
cyclic deformation of the configuration, as defined
in eq.~(\ref{eq:3cycle}), with no extra overall
rotation of the triangle. It gives a phase factor
$\rme^{2\srmi\theta}$ in the wave function, where
$\theta=\nu\pi$ is the statistics parameter. If we deform
while keeping all the time $|u|<|v|$, the phase of
$v$ increases continuously from $\varphi_2$ to
$\varphi_2+(2\pi/3)$, whereas the phase of $u$
changes from $\varphi_1$ to
$\varphi_1-(2\pi/3)+2m'\pi$, where $m'$
is any integer. The corresponding boundary
condition on the wave function is, therefore,
\beqn
\psi\!\left(r,q,\varphi_1-{2\pi\over 3}+2m'\pi\,,\,
\varphi_2+{2\pi\over 3}\right)=
\rme^{\,2\srmi\theta}\,
\psi(r,q,\varphi_1,\varphi_2)\;.
\eeqn
That is,
\beqn
j\left(-{2\pi\over 3}+2m'\pi\right)
+k\,{2\pi\over 3}=2(n'+\nu)\pi\;,
\eeqn
for some integer $n'$. Since $m'$ is an arbitrary
integer, $j$ must be an integer. Then
\beqn
k=j+3(n+\nu)\;,
\eeqn
where $n=n'-jm'$ is an arbitrary integer, and
the eigenvalue of the relative angular momentum
$\Lrel$ is
\beqn
\label{eq:Lrelqu}
\ell=j+k=2j+3(n+\nu)\;.
\eeqn

These relations take care of the cyclic
permutations of all three particles. What
remains is only to take care of
one of the three cases where two particles
are interchanged, for example
$z_2\leftrightarrow z_3$, or equivalently,
$u\leftrightarrow v$. This is the same as
$q\leftrightarrow 1/q$ and
$\varphi_1\leftrightarrow\varphi_2$, if
we define angles so that $u=v$ corresponds
to $\varphi_1=\varphi_2$.
To be more precise, we consider a continuous
interchange, with $q=1$ at the beginning and
end, and $q<1$ during the interchange.
The interchange should be anticlockwise,
which means that we start with
$\varphi_1>\varphi_2$, and end with
$\varphi_1<\varphi_2$. There is one
further restriction, that
$|\varphi_1-\varphi_2|<(2\pi/3)$ when $q=1$,
meaning that the particle
position $z_1$ must not be encircled.

Thus, the boundary condition on $\psi$
at $q=1$ is
\beqn
\psi(r,1,\varphi_2,\varphi_1)=
\rme^{\,\srmi\theta}\,
\psi(r,1,\varphi_1,\varphi_2)\;,
\eeqn
for $0<\varphi_1-\varphi_2<(2\pi/3)$.
However, this condition is incomplete,
because the general condition is
\beqn
\psi(r,1/q,\varphi_2,\varphi_1)=
\rme^{\,\srmi\theta}\,
\psi(r,q,\varphi_1,\varphi_2)\;.
\eeqn
Since the Schr{\"o}dinger equation is second
order in the $q$ derivative, we need boundary
conditions at $q=1$ both for
the wave function $\psi$ and its normal
derivative $\psi_q=\partial\psi/\partial q$.
The derivative condition is easily deduced,
\beqn
\psi_q(r,1,\varphi_2,\varphi_1)=
-\rme^{\,\srmi\theta}\,
\psi_q(r,1,\varphi_1,\varphi_2)\;.
\eeqn

The boundary conditions for $\psi$ and $\psi_q$
can not be satisfied by a wave function which is
separable in $q$, $\varphi_1$ and $\varphi_2$.
But we may quantize the relative angular
momentum $\ell$, and according to
eq.~(\ref{eq:Lrelqu}) $\ell-3\nu=2j+3n$ is
an integer, either even or odd.
Let $\nu'=\nu$ if $n=2m$ and $\nu'=\nu+1$
if $n=2m+1$, with $m$ integer. Then
\beqn
j={\ell\over 2}
-3\left(m+{\nu'\over 2}\right),\qquad
k={\ell\over 2}
+3\left(m+{\nu'\over 2}\right).
\eeqn
Let $g_m(q)$ be the function $g(q)$ as
given by eq.~(\ref{eq:gqdef}).
Introducing $\varphi=(\varphi_1+\varphi_2)/2$,
and summing over $m$, including an as yet
undetermined coefficient $C_m$ for each $m$,
we get the following angular wave function,
\beqn
\Omega(q,\varphi_1,\varphi_2)=
\sum_{m=-\infty}^{\infty}
C_m\,g_m(q)\,\rme^{\,\srmi\,(j\varphi_1+k\varphi_2)}
=\rme^{\,\srmi\,\ell\varphi}
\sum_{m=-\infty}^{\infty}
C_m\,g_m(q)\,\rme^{\,-3\srmi(m+(\nu'/2))
(\varphi_1-\varphi_2)}\;.
\eeqn
It is natural to call $\Omega$
an {\em anyonic spherical harmonic function},
whenever it satisfies the anyonic boundary
conditions.

Define $\xi=3(\varphi_1-\varphi_2)$.
The two boundary conditions that must
hold for $0<\xi<2\pi$ are
\beqn
\label{eq:boundaryconds}
\sum_{m=-\infty}^{\infty}
C_m\,g_m(1)\,\rme^{\,\srmi m\xi}
&\!\!\!=&\!\!\!
\;\,\,\rme^{\,\srmi(\nu\pi-\nu'\xi)}
\sum_{m=-\infty}^{\infty}
C_m\,g_m(1)\,\rme^{\,-\srmi m\xi}\;,
\nonumber\\
\sum_{m=-\infty}^{\infty}
C_m\,g'_m(1)\,\rme^{\,\srmi m\xi}
&\!\!\!=&\!\!\!
-\rme^{\,\srmi(\nu\pi-\nu'\xi)}
\sum_{m=-\infty}^{\infty}
C_m\,g'_m(1)\,\rme^{\,-\srmi m\xi}\;.
\eeqn

Recall that $g_m(1)$ and $g'_m(1)$ depend on
the three parameters
$\mu=E-2-2n_r$, $\ell$ and $m$.
We claim that for each given $\ell$,
the parameter $\mu$, which determines
the energy $E$, may be adjusted so that
the above boundary conditions have non-trivial
solutions for the coefficients $C_m$.
For each $\ell$ there will be many solutions,
possibly many with the same $\mu$, and this
procedure should give the complete
set of anyonic spherical harmonics.
A complete proof would involve a study
of the anyonic spherical harmonics, which
we have not yet done, but we can give
a plausibility argument, which also
indicates one possible way to solve
the problem numerically.

We may regard $\gamma_m=C_m\,g_m(1)$ and
$\wt{\gamma}_m=C_m\,g'_m(1)$ as the Fourier
components of two functions
\beqn
\gamma(\xi)&\!\!\!=&\!\!\!
\sum_{m=-\infty}^{\infty}
C_m\,g_m(1)\,\rme^{\,\srmi m\xi}\;,
\nonumber\\
\wt{\gamma}(\xi)&\!\!\!=&\!\!\!
\sum_{m=-\infty}^{\infty}
C_m\,g'_m(1)\,\rme^{\,\srmi m\xi}\;,
\eeqn
periodic in $\xi$ with period $2\pi$.
We may restrict their regions of definition
to be the interval ${[0,2\pi]}$.

There is a natural scalar product
between any two functions $\phi=\phi(\xi)$
and $\chi=\chi(\xi)$, with Fourier components
$\phi_m$ and $\chi_m$ as above,
\beqn
(\phi,\chi)=\int_0^{2\pi}\rmd\xi\,
(\phi(\xi))^{\ast}\,\chi(\xi)
=2\pi\sum_{m=-\infty}^{\infty}
(\phi_m)^{\ast}\,\chi_m\;.
\eeqn

Define the linear operator $A$ by
\beqn
{[A\phi]}(\xi)=
\rme^{\srmi(\nu\pi-\nu'\xi)}\,
\phi(2\pi-\xi)\;,
\eeqn
for $0<\xi<2\pi$.
Then $A$ is Hermitean with respect to
the natural scalar product, and $A^2=I$,
the identity operator.
Note that $A$ is a somewhat singular
operator, unless $\nu$ is an integer,
since the factor $\rme^{\srmi(\nu\pi-\nu'\xi)}$,
extended by periodicity outside the interval
${[0,2\pi]}$, is discontinuous at every
integer multiple of $2\pi$.

Define another operator $B$ such that
$B\gamma=\wt{\gamma}$. It is also Hermitean, since
it is diagonalized by the Fourier
transformation, and its eigenvalues
$\beta_m=g'_m(1)/g_m(1)$ are real.
The two boundary conditions are simply
$\gamma=A\gamma$, $B\gamma=-AB\gamma$,
equivalent to two simultaneous eigenvalue
equations,
\beqn
A\gamma=\gamma\;,\qquad
(I+A)B(I+A)\gamma=0\;.
\eeqn
These two equations are compatible, since
the operators $A$ and $B_+=(I+A)B(I+A)$
commute, and $B_+$ has a complete set of
real eigenvalues, since it is Hermitean.
If we select one eigenvalue of $B_+$ on
the subspace where $A\gamma=\gamma$,
and set it equal to zero,
this is one real equation for one real
parameter $\mu$. By equating different
eigenvalues to zero we should obtain
a complete set of solutions.

There is one minor problem with this argument
in that $B$ is undefined if an eigenvalue
$\beta_m=g'_m(1)/g_m(1)$ is infinite.
There is of course a similar argument using
$B^{-1}$ instead of $B$, but in principle
it might happen that $B$ and $B^{-1}$ are
undefined simultaneously.

\section{Numerical Solution}

To find a numerical solution for the coefficients
$C_m$ satisfying the boundary conditions in
eq.~(\ref{eq:boundaryconds}), we
must truncate to a finite number $N$ of coefficients,
and in order to use the fast Fourier transform,
we choose $N$ as a power of 2. Then we impose
eq.~(\ref{eq:boundaryconds}) at the $N$ discrete points
\beqn
\xi_k={(2k-1)\pi\over N}\;,\qquad\mbox{for}\quad
k=1,2,\ldots,N\;.
\eeqn

It is a remarkable feature of our method that for
very small $N$, say $N=4$, it can give many energy
levels with good accuracy. This is so when the low
Fourier components, i.e.\ with small $m$, dominate.
On the other hand, the convergence as $N\to\infty$
is sometimes very slow. This is clearly related
to the fact that the wave functions for non-integer
$\nu$ have non-integer power behaviour at $\xi=0$,
where two particles meet. Hence our approximations
by finite Fourier series converge slowly. Note that
eq.~(\ref{eq:boundaryconds}) at $\xi=0$ gives that
\beqn
(1-\rme^{\,\srmi\nu\pi})\,\gamma(0)=
(1+\rme^{\,\srmi\nu\pi})\,\wt{\gamma}(0)=0\;,
\eeqn
assuming that the wave function is continuous.
Thus $\gamma(0)=0$, except in the boson case, and
$\wt{\gamma}(0)=0$, except in the fermion case.
For our numerical solutions these conditions at
$\xi=0$ are not imposed and will hold only in
the asymptotic limit $N\to\infty$.

To illustrate the convergence we have tabulated
the approximate energy $E_N$ as a function of $N$,
in three cases. Table 1 is for the state which
becomes the bosonic ground state at $\nu=0$.
It is exactly known and has energy $E=2+3\nu$
and angular momentum $\ell=3\nu$.
Table 2 is for the state which
becomes the fermionic ground state at $\nu=1$.
It has angular momentum $\ell=-3+3\nu$ and
an energy depending on $\nu$ in an unknown way.
Table 3 is for the state connected to the fermionic
ground state by the supersymmetry transformation of
Sen \cite{Sen92,Sen92a}. It has angular momentum
$\ell=-2+3\nu$. The energy eigenvalues in Tables 2
and 3 should be the same, except that $\nu$ in one
table corresponds to $1-\nu$ in the other.

It appears from the tables that the leading correction
term for finite $N$ is of order $N^{-2\nu}$. Using
two different $N$ one may therefore extrapolate
to $N=\infty$, and this improves the convergence
considerably. Another point to note is that one may
take advantage of the supersymmetry in order
to get more accurate energy levels.

\section*{Acknowledgments}

Many people have shared with us their insight.
We want to thank especially
Jon {\AA}ge Ruud and Finn Ravndal, St{\'e}phane
Ouvry who gave us an unpublished report
by Pierre Baumann, and Matts Sporre and collaborators
who gave us access to their numerical data.


\begin{table}[p]
\caption{The bosonic ground state energy}
\begin{center}
\begin{tabular}{|l|l|l|l|l|}
\cline{2-5}
\multicolumn{1}{c|}{} & \multicolumn{1}{c|}{$\nu = 0.2$} &
\multicolumn{1}{c|}{$\nu = 0.4$} & \multicolumn{1}{c|}{$\nu = 0.6$} &
\multicolumn{1}{c|}{$\nu = 0.8$} \\
\hline
\multicolumn{1}{|r|}{$E_{\mbox{\scriptsize exact}}$}
 & 2.6 & 3.2 & 3.8 & 4.4 \\
\hline
$N=2$  &2.3279 &2.972\,08 &3.666\,068\,15 &4.345\,506\,34\\
$N=4$  &2.3786 &3.063\,17 &3.743\,586\,44 &4.384\,285\,82\\
$N=8$  &2.4228 &3.119\,16 &3.775\,664\,77 &4.395\,001\,04\\
$N=16$ &2.4599 &3.152\,74 &3.789\,423\,68 &4.398\,366\,91\\
$N=32$ &2.4902 &3.172\,56 &3.795\,395\,68 &4.399\,462\,67\\
$N=64$ &2.5146 &3.184\,14 &3.797\,995\,23 &4.399\,822\,87\\
\hline
\multicolumn{5}{|l|}{$\gamma$ from fit to
  $E_N = E_{\mbox{\scriptsize exact}} + A N^{-\gamma}$}\\
\hline
$N=2$, $4$   & 0.298 & 0.7362 & 1.247\,39 & 1.794\,02\\
$N=4$, $8$   & 0.321 & 0.7592 & 1.213\,00 & 1.652\,37\\
$N=8$, $16$  & 0.339 & 0.7745 & 1.202\,21 & 1.614\,02\\
$N=16$, $32$ & 0.352 & 0.7845 & 1.199\,78 & 1.603\,72\\
$N=32$, $64$ & 0.362 & 0.7908 & 1.199\,56 & 1.600\,98\\
\hline
\multicolumn{5}{|l|}{$E_\infty$ from fit to
  $ E_N = E_\infty + A N^{-2\nu}$}\\
\hline
$N=2$, $4$   &2.5375 &3.186\,09 &3.803\,335\,54 &4.403\,375\,54\\
$N=4$, $8$   &2.5611 &3.194\,70 &3.800\,389\,92 &4.400\,275\,74\\
$N=8$, $16$  &2.5759 &3.198\,05 &3.800\,028\,70 &4.400\,023\,81\\
$N=16$, $32$ &2.5851 &3.199\,31 &3.799\,998\,73 &4.400\,002\,07\\
$N=32$, $64$ &2.5909 &3.199\,76 &3.799\,998\,91 &4.400\,000\,18\\
\hline
\end{tabular}
\end{center}
\end{table}

\begin{table}[p]
\caption{The fermionic ground state energy}
\begin{center}
\begin{tabular}{|l|l|l|l|l|}
\cline{2-5}
\multicolumn{1}{c|}{} & \multicolumn{1}{c|}{$\nu = 0.2$} &
\multicolumn{1}{c|}{$\nu = 0.4$} & \multicolumn{1}{c|}{$\nu = 0.6$} &
\multicolumn{1}{c|}{$\nu = 0.8$} \\
\hline
$N=2$   &4.6455 &4.380\,322 &4.181\,3475 &4.048\,5533\\
$N=4$   &4.6356 &4.379\,612 &4.181\,8967 &4.047\,9552\\
$N=8$   &4.6446 &4.394\,153 &4.189\,5454 &4.049\,7399\\
$N=16$  &4.6552 &4.405\,332 &4.194\,0494 &4.050\,6380\\
$N=32$  &4.6646 &4.412\,454 &4.196\,2485 &4.050\,9943\\
$N=64$  &4.6723 &4.416\,732 &4.197\,2586 &4.051\,1243\\
$N=128$ &4.6786 &4.419\,241 &4.197\,7105 &4.051\,1700\\
\hline
\multicolumn{5}{|l|}{$\gamma$ from fit to
  $ E_N = E_\infty + A N^{-\gamma}$} \\
\hline
$N=8$, $16$, $32$   &0.171 &0.650 &1.034 &1.334 \\
$N=16$, $32$, $64$  &0.276 &0.735 &1.122 &1.454 \\
$N=32$, $64$, $128$ &0.319 &0.770 &1.160 &1.512 \\
\hline
\multicolumn{5}{|l|}{$E_\infty$ from fit to
  $ E_N = E_\infty + A N^{-\gamma}$} \\
\hline
$N=8$, $16$, $32$   &4.7394 &4.424\,958 &4.198\,3467 &4.051\,2286\\
$N=16$, $32$, $64$  &4.7092 &4.423\,167 &4.198\,1166 &4.051\,1991\\
$N=32$, $64$, $128$ &4.7038 &4.422\,799 &4.198\,0764 &4.051\,1946\\
\hline
\multicolumn{5}{|l|}{$E_\infty$ from fit to
  $ E_N = E_\infty + AN^{-2\nu}$}\\
\hline
$N=2$, $4$    &4.6046 &4.378\,652 &4.182\,3200 &4.047\,6609\\
$N=4$, $8$    &4.6726 &4.413\,775 &4.195\,4408 &4.050\,6184\\
$N=8$, $16$   &4.6883 &4.420\,416 &4.197\,5209 &4.051\,0800\\
$N=16$, $32$  &4.6940 &4.422\,065 &4.197\,9435 &4.051\,1697\\
$N=32$, $64$  &4.6966 &4.422\,505 &4.198\,0371 &4.051\,1884\\
$N=64$, $128$ &4.6981 &4.422\,582 &4.198\,0588 &4.051\,1924\\
\hline
\end{tabular}
\end{center}
\end{table}

\begin{table}[p]
\caption{The supersymmetric partner of
the fermionic ground state}
\begin{center}
\begin{tabular}{|l|l|l|l|l|}
\cline{2-5}
\multicolumn{1}{c|}{} & \multicolumn{1}{c|}{$\nu = 0.2$} &
\multicolumn{1}{c|}{$\nu = 0.4$} & \multicolumn{1}{c|}{$\nu = 0.6$} &
\multicolumn{1}{c|}{$\nu = 0.8$} \\
\hline
$N=2$  &3.8436 &3.954\,56 &4.249\,4885 &4.627\,147\,52\\
$N=4$  &3.8814 &4.051\,14 &4.348\,6537 &4.678\,013\,75\\
$N=8$  &3.9144 &4.111\,77 &4.391\,1581 &4.692\,955\,72\\
$N=16$ &3.9423 &4.147\,84 &4.409\,1615 &4.697\,703\,46\\
$N=32$ &3.9653 &4.168\,98 &4.416\,8475 &4.699\,244\,71\\
$N=64$ &3.9840 &4.181\,27 &4.420\,1526 &4.699\,748\,11\\
\hline
\multicolumn{5}{|l|}{$\gamma$ from fit to
  $ E_N = E_\infty + A N^{-\gamma}$} \\
\hline
$N=2$, $4$, $8$    &0.194 &0.672 &1.222 &1.767\\
$N=4$, $8$, $16$   &0.246 &0.749 &1.239 &1.654\\
$N=8$, $16$, $32$  &0.275 &0.772 &1.228 &1.623\\
$N=16$, $32$, $64$ &0.299 &0.782 &1.218 &1.614\\
\hline
\multicolumn{5}{|l|}{$E_\infty$ from fit to
  $ E_N = E_\infty + A N^{-\gamma}$} \\
\hline
$N=2$, $4$, $8$    &4.1441 &4.214\,00 &4.423\,0430 &4.699\,170\,53\\
$N=4$, $8$, $16$   &4.0925 &4.200\,86 &4.422\,3905 &4.699\,914\,63\\
$N=8$, $16$, $32$  &4.0749 &4.198\,87 &4.422\,5732 &4.699\,985\,54\\
$N=16$, $32$, $64$ &4.0652 &4.198\,36 &4.422\,6461 &4.699\,992\,28\\
\hline
\multicolumn{5}{|l|}{$E_\infty$ from fit to
  $ E_N = E_\infty + AN^{-2\nu}$}\\
\hline
$N=2$, $4$   &3.9997 &4.181\,46 &4.425\,0877 &4.703\,053\,33 \\
$N=4$, $8$   &4.0178 &4.193\,57 &4.423\,9194 &4.700\,311\,10 \\
$N=8$, $16$  &4.0295 &4.196\,52 &4.423\,0381 &4.700\,040\,61 \\
$N=16$, $32$ &4.0374 &4.197\,49 &4.422\,7717 &4.700\,003\,41 \\
$N=32$, $64$ &4.0426 &4.197\,86 &4.422\,7001 &4.699\,995\,92 \\
\hline
\end{tabular}
\end{center}
\end{table}

\end{document}